# Bound states and the potential parameter spectrum

A. D. Alhaidari[a] and H. Bahlouli[b]

[a] *Saudi Center for Theoretical Physics, P.O. Box 32741, Jeddah 21438, Saudi Arabia*

[b] *Physics Department, King Fahd University of Petroleum & Minerals, Dhahran 31261, Saudi Arabia*

**Abstract**: In this article, we answer the following question: If the wave equation possesses bound states but it is exactly solvable for only a single non-zero energy, can we find all bound state solutions (energy spectrum and associated wavefunctions)? To answer this question, we use the "tridiagonal representation approach" to solve the wave equation at the given energy by expanding the wavefunction in a series of energy-dependent square integrable basis functions in configuration space. The expansion coefficients satisfy a three-term recursion relation, which is solved in terms of orthogonal polynomials. Depending on the selected energy we show that one of the potential parameters must assume a value from within a discrete set called the "potential parameter spectrum" (PPS). This discrete set is obtained from the spectrum of the above polynomials and can be either a finite or an infinite discrete set. Inverting the relation between the energy and the PPS gives the bound state energy spectrum. Therefore, the answer to the above question is affirmative.

## 1. Introduction

In the past, we have encountered situations in which the wave equation in quantum mechanics is endowed with bound states but it is exactly solvable for only a single non-zero energy. Changing the value of one of the potential parameters to another value from within a specific discrete set leads to a different exact solution but for the same energy [1,2]. However, this is not an energy degenerate solution because the potential function is not the same. We call this discrete set of parameter values the "potential parameter spectrum" (PPS), which could be finite or infinite in size. The PPS concept was introduced for the first time in [3] while searching for quasi-exact solutions of the wave equation. In this work, we use the Tridiagonal Representation Approach (TRA) [4] to make a systematic study of this problem with illustrative examples. In the TRA, the quantum mechanical wave function is written as a linear sum (finite or infinite) of a complete set of square integrable basis functions in configuration space. The elements of the basis set are required to be compatible with the boundary conditions in configuration space and should produce a tridiagonal matrix representation for the wave operator $J = H - E$, where $H$ is the Hamiltonian of the system and $E$ is its energy. Consequently, the matrix wave equation becomes a three-term recursion relation for the expansion coefficients of the wavefunction. The solution of this recursion is then expressed in terms of orthogonal polynomials that depend on the energy and the potential parameters of the problem. All physical properties of the system (e.g., energy spectrum, phase shift, density of states, etc.) are obtained from the properties of these polynomials (e.g., asymptotics, zeros, weight function, etc.). In particular, and as relevant to the current study, we use the spectrum formula of these polynomials to give the PPS as a function of the energy.

Inverting this relation gives the bound states energy spectrum. As such, we are able to deduce the full bound state solution of the problem from its solution at a single energy.

We write the Hamiltonian operator as $H = T + V$ where $T$ is the kinetic energy operator and $V$ is the potential function and we are interested in solving the stationary Schrödinger equation, $H|\psi(E)\rangle = E|\psi(E)\rangle$. For this purpose, we let $\{\phi_n(x)\}$ be a complete $L^2$ basis set with $x$ being the configuration space coordinate and write the wavefunction as $|\psi(E)\rangle = \sum_n f_n(E)|\phi_n\rangle$, where $\{f_n\}$ is a proper set of expansion coefficients. The TRA requirements dictate that $\{f_n\}$ satisfy a three-term recursion relation resulting from the matrix wave equation. Since the solution of any three-term recursion relation is defined up to an overall factor, we define $f_n = f_0 P_n$ making $P_0 = 1$, then $\{P_n\}$ will be a set of orthogonal polynomials whose positive definite weight function is $f_0^2$ [5,6]. If we denote the matrix elements of an operator $R$ in the basis by $R_{n,m} = \langle \phi_n | R | \phi_m \rangle$ then $J_{n,m} = T_{n,m} + V_{n,m} - E\Omega_{n,m}$, where $\Omega_{n,m} = \langle \phi_n | \phi_m \rangle$ is the basis overlap matrix (the matrix representation of the identity). Now, $T$ is usually a well-known differential operator in configuration space. For example, in one dimension with coordinate $x$ and adopting the atomic units $\hbar = m = 1$, $T = -\frac{1}{2}\frac{d^2}{dx^2}$. In three dimensions with spherical symmetry and radial coordinate $r$, $T = -\frac{1}{2}\frac{d^2}{dr^2} + \frac{\ell(\ell+1)}{2r^2}$ where $\ell$ is the angular momentum quantum number. Therefore, the action of $T$ on the given basis elements $\{\phi_n(x)\}$ could easily be derived and so too its matrix elements which will certainly depend on the basis parameters.

If $\Omega$ turns out to be a tridiagonal matrix (including, of course, the special case of a diagonal matrix for orthogonal basis) then the wave operator matrix $J$ can be made tridiagonal by requiring that any non-tridiagonal components in $T$ be eliminated by counter terms in $V$. Consequently, the basis parameters in the TRA solution of such a problem will depend on the potential parameters and there will be no constraint on the value of the energy to obtain the solution. However, for the purpose of the current study we impose the condition that $\Omega$ is non-tridiagonal so that the energy will be fixed by the tridiagonalization constraint. Therefore, $\Omega$ could be penta-diagonal (5-term banded diagonal), hepta-diagonal (7-term banded diagonal), etc. In general, it has a $(2k+1)$-term diagonal band where $k = 2, 3, 4...$ In fact, it could even be a full matrix. With such an $\Omega$, the matrix $J$ could still be made tridiagonal while maintaining an energy independent potential. In such a scenario, the kinetic energy matrix $T$ must be non-tridiagonal such that its non-tridiagonal component cancels $E\Omega$. That is, if we write $T_{n,m} = T_{n,m}^0 + \tilde{T}_{n,m}$ where $\tilde{T}_{n,m}$ is the non-tridiagonal component, then we choose the basis parameters such that $\tilde{T}_{n,m} = E\Omega_{n,m}$ making $J_{n,m} = T_{n,m}^0 + V_{n,m}$ where $V$ is tridiagonal. The equation $\tilde{T}_{n,m} = E\Omega_{n,m}$ dictates that the basis parameters in this TRA solution be energy dependent. In some cases, however, the matrix $T$ contains an extra non-tridiagonal component beyond that which is used to eliminate $E\Omega$. In such cases, if we write $T_{n,m} = T_{n,m}^0 + \tilde{T}_{n,m} + \bar{T}_{n,m}$ then the potential matrix must be non-tridiagonal as $V_{n,m} = V_{n,m}^0 + \bar{V}_{n,m}$ and we require that $\bar{V}_{n,m} = -\bar{T}_{n,m}$ and $\tilde{T}_{n,m} = E\Omega_{n,m}$ resulting in a tridiagonal matrix $J$ with the elements $J_{n,m} = T_{n,m}^0 + V_{n,m}^0$. In this case, the basis parameters are not only energy dependent but they also depend on the potential parameters. In rare cases, however, no non-tridiagonal component of $T$ could be used to eliminate the non-tridiagonal matrix $E\Omega$. For such cases, the problem is solvable only at zero energy and no PPS exists.

In the following section, we formulate the problem using an illustrative example in 3D. We identify the orthogonal polynomial from which we obtain the PPS formula and invert it to get the energy spectrum for the system. In section 3 and 4, we give two other examples in 1D. Conclusion and discussion are presented in section 5.

## 2. TRA formulation of the problem: The Kratzer potential

We consider a problem in three dimensions with spherical symmetry and take the following radial functions as elements of the complete basis set

$$\phi_n(r) = (\lambda r)^\mu e^{-\lambda r/2} L_n^\nu(\lambda r), \tag{1}$$

where $L_n^\nu(x)$ is the Laguerre polynomial, $\lambda$ is a positive scale parameter having dimension of inverse length and $\{\mu,\nu\}$ are dimensionless real parameters such that $\nu > -1$. For a tridiagonal overlap matrix $\Omega$, the TRA solution has already been worked out where the corresponding problem turned out to be the Coulomb problem [7]. Here, we consider the case where $\Omega$ is non-tridiagonal. Now, in the atomic units $\hbar = m = 1$, the action of the wave operator on the basis element (1) is written as follows

$$-\frac{2}{\lambda^2} J |\phi_n\rangle = \left[ \frac{d^2}{dx^2} - \frac{\ell(\ell+1)}{x^2} - \frac{2V(x)}{\lambda^2} + \frac{2E}{\lambda^2} \right] x^\mu e^{-x/2} L_n^\nu(x)$$

$$= x^\mu e^{-x/2} \left[ \frac{d^2}{dx^2} + \left( \frac{2\mu}{x} - 1 \right) \frac{d}{dx} + \frac{\mu(\mu-1) - \ell(\ell+1)}{x^2} - \frac{\mu}{x} + \frac{1}{4} - \frac{2V(x)}{\lambda^2} + \frac{2E}{\lambda^2} \right] L_n^\nu(x) \tag{2}$$

where $x = \lambda r$. Employing the differential equation of the Laguerre polynomials [8-10], this equation becomes

$$-\frac{2}{\lambda^2} J |\phi_n\rangle = x^\mu e^{-x/2} \left[ \frac{2\mu - \nu - 1}{x} \frac{d}{dx} + \frac{\left(\mu - \frac{1}{2}\right)^2 - \left(\ell + \frac{1}{2}\right)^2}{x^2} - \frac{n+\mu}{x} + \frac{1}{4} - \frac{2V}{\lambda^2} + \frac{2E}{\lambda^2} \right] L_n^\nu. \tag{3}$$

The differential relation of the Laguerre polynomial, $x \frac{d}{dx} L_n^\nu = n L_n^\nu - (n+\nu) L_{n-1}^\nu$, maps this equation into the following

$$-\frac{2}{\lambda^2} J |\phi_n\rangle = x^\mu e^{-x/2}$$

$$\left\{ \left[ \frac{\left(\mu - \frac{1}{2}\right)^2 - \left(\ell + \frac{1}{2}\right)^2 + (2\mu - \nu - 1)n}{x^2} - \frac{n+\mu}{x} + \frac{1}{4} - \frac{2V}{\lambda^2} + \frac{2E}{\lambda^2} \right] L_n^\nu - \frac{2\mu - \nu - 1}{x^2} (n+\nu) L_{n-1}^\nu \right\} \tag{4}$$

Now, $\Omega_{n,m} = \langle \phi_n | \phi_m \rangle = \int_0^\infty x^{2\mu} e^{-x} L_n^\nu(x) L_m^\nu(x) dx$ should not be tridiagonal. Thus, the orthogonality of the Laguerre polynomials and its recursion relation [8-10] show that the parameter choice $2\mu = \nu + 1$ that could have eliminated the $L_{n-1}^\nu$ term in Eq. (4) cannot be

−3−

made since it will result in a tridiagonal matrix $\Omega$. For the same reason we cannot even make the choice $2\mu = \nu$ that would have made the basis orthogonal and $\Omega$ a diagonal matrix. On the other hand, factoring out $1/x^2$ from the curly brackets, we obtain

$$-\frac{2}{\lambda^2}J|\phi_n\rangle = x^{\mu-2}e^{-x/2}$$
$$\left\{\left[\left(\mu-\tfrac{1}{2}\right)^2 - \left(\ell+\tfrac{1}{2}\right)^2 + (2\mu-\nu-1)n - (n+\mu)x + \frac{x^2}{4} + \frac{2x^2}{\lambda^2}(E-V)\right]L_n^\nu - (2\mu-\nu-1)(n+\nu)L_{n-1}^\nu\right\} \quad (5)$$

The recursion relation and orthogonality of the Laguerre polynomials show that the matrix representation of the wave operator $\langle\phi_m|J|\phi_n\rangle$ is tridiagonal if and only if $2\mu = \nu + 2$ and the expression inside the square brackets in (5) is linear in $x$. Therefore, the basis overlap matrix $\Omega$ becomes penta-diagonal and to obtain a linear expression in $x$ with an energy independent potential, we impose the following constraints

$$\lambda^2 = -8E, \quad (6a)$$

$$V(x) = \frac{\lambda^2}{2}\frac{\alpha x + \beta}{x^2} = \frac{\alpha\lambda}{2r} + \frac{\beta}{r^2}, \quad (6b)$$

where $\alpha$ and $\beta$ are arbitrary dimensionless real parameters. This potential is the Kratzer potential [11], which is the sum of the Coulomb (where the electric charge is $Z = \alpha\lambda/2$) and an inverse square potential of strength $\beta$. This case corresponds to $J_{n,m} = T_{n,m}^0 + V_{n,m}$ where the non-tridiagonal matrix $E\Omega$ has been eliminated by the non-tridiagonal component of the kinetic energy matrix $\tilde{T}$ using the special choice of the basis parameter (6a). Reality of the parameter $\lambda$ in Eq. (6a) dictates that the energy must be negative corresponding to bound states. Substituting (6) along with $2\mu = \nu + 2$ in Eq. (5), we obtain

$$-\frac{2}{\lambda^2}J|\phi_n\rangle = x^{\frac{\nu}{2}-1}e^{-x/2}\left\{\left[\tfrac{1}{4}(\nu+1)^2 - \left(\ell+\tfrac{1}{2}\right)^2 - \beta + n - \left(n+\tfrac{\nu}{2}+\alpha+1\right)x\right]L_n^\nu - (n+\nu)L_{n-1}^\nu\right\} \quad (7)$$

Using the three-term recursion relation of the Laguerre polynomials, $xL_n^\nu = (2n+\nu+1)L_n^\nu - (n+\nu)L_{n-1}^\nu - (n+1)L_{n+1}^\nu$, this equation becomes

$$-\frac{2}{\lambda^2}J|\phi_n\rangle = x^{\frac{\nu}{2}-1}e^{-x/2}\left\{-\left[(2n+\nu+1)\left(n+\tfrac{\nu}{2}+\alpha+1\right) - n + \beta + \left(\ell+\tfrac{1}{2}\right)^2 - \tfrac{1}{4}(\nu+1)^2\right]L_n^\nu \right.$$
$$\left. + (n+\nu)\left(n+\tfrac{\nu}{2}+\alpha\right)L_{n-1}^\nu + (n+1)\left(n+\tfrac{\nu}{2}+\alpha+1\right)L_{n+1}^\nu\right\} \quad (8)$$

Substituting $|\psi\rangle = \sum_n f_n|\phi_n\rangle$ in the wave equation $J|\psi\rangle = 0$ and using Eq. (8), we obtain the following three-term recursion relation for the expansion coefficients

$$-\left[2n(n+\nu+\alpha+1) + (\nu+1)\left(\tfrac{\nu}{2}+\alpha+1\right) + \left(\ell+\tfrac{1}{2}\right)^2 - \tfrac{1}{4}(\nu+1)^2 + \beta\right]P_n$$
$$+ n\left(n+\tfrac{\nu}{2}+\alpha\right)P_{n-1} + (n+\nu+1)\left(n+\tfrac{\nu}{2}+\alpha+1\right)P_{n+1} = 0 \quad (9)$$

where we wrote $f_n = f_0 P_n$. Comparing this recursion relation to that of the continuous dual Hahn polynomial shown as Eq. (A3) in Appendix A, we conclude that $P_n = S_n(z^2;a,b,c)$ where

$$a = b = \tfrac{1}{2}(\nu+1), \qquad c = \alpha + \tfrac{1}{2}, \qquad z^2 = -\left(\ell+\tfrac{1}{2}\right)^2 - \beta. \qquad (10)$$

Now, the spectrum of the polynomial $S_n(z^2;a,b,c)$ is either continuous or discrete depending on the value of $z^2$ [12]. If $\beta \leq -\left(\ell+\tfrac{1}{2}\right)^2$ then $z^2 \geq 0$ and the spectrum is continuous. However, if $\beta > -\left(\ell+\tfrac{1}{2}\right)^2$ then $z^2 < 0$ and the spectrum is discrete. Let us now make a physical investigation of the problem and see which one of the two cases apply. On physical grounds, it is well known that the inverse square potential with coupling strength less than the critical value of $-1/8$ is plagued with quantum anomalies. Most prominent of these is the rapid and unbounded increase in oscillations of the wave function as the particle nears the origin. Hence, the energy spectrum is abnormal because it is not bounded from below assuming all negative values that extend to minus infinity [13,14]. Landau and Lifshitz associated the occurrence of these infinite bound states to the classical "particle fall to the center" [15,16]. These anomalies could be resolved using various methods of regularization, most of them without unique results [13,14,17]. In our problem, the inverse square component of the effective potential is $\tfrac{1}{2}[\ell(\ell+1)+\beta]r^{-2}$. Thus, the super-critical coupling regime corresponds to values of the coupling parameter $\beta \leq -\left(\ell+\tfrac{1}{2}\right)^2$. Hence, we discard this choice and consider the case where $\beta > -\left(\ell+\tfrac{1}{2}\right)^2$. Moreover, since we are interested in bound state solutions, then the wavefunction must vanish at infinity. That is, $\lim_{r\to\infty}\psi(r) = 0$. However, it has been established elsewhere that if $\psi(r) = \sum_{n=0}^{\infty} f_n \phi_n(r)$ then $\lim_{r\to\infty}\psi(r) \propto \lim_{N\to\infty}\sum_{n=N}^{\infty} f_n \phi_n(r)$ [18]. Consequently, the asymptotic ($n \to \infty$) of $f_n$ (equivalently, the asymptotics of $P_n$) must vanish. Now equation (A4) in Appendix A shows that the asymptotics of $S_n(z^2;a,b,c)$ is sinusoidal with the following amplitude

$$A(z) = \frac{2\Gamma(b+c)|\Gamma(2iz)|n^{-a}}{|\Gamma(b+iz)\Gamma(c+iz)\Gamma(a+iz)|}. \qquad (11)$$

This amplitude can be made to vanish if any argument of the three gamma functions in the denominator becomes a negative integer or zero. If $\gamma$ stands for any one of the parameters $\{a,b,c\}$, then $\gamma + iz = -k$ where $k = 0,1,...,N$ with $N$ being the largest integer less than or equal to $-\gamma$. Thus, $\gamma$ must be negative or zero. Equation (10) shows that both $a$ and $b$ are positive since $\nu > -1$ and hence $\gamma = c$. With the values of $c$ shown in Eq. (10), we require that $\alpha \leq -\tfrac{1}{2}$ (i.e., $Z \leq -\tfrac{\lambda}{4}$) and $\gamma + iz = -k$ giving $z^2 = -(k+c)^2$ and resulting in the following relation

$$\beta = \left(k+\ell+1+\tfrac{2Z}{\lambda}\right)\left(k-\ell+\tfrac{2Z}{\lambda}\right), \qquad (12)$$

where $k = 0,1,...,N$ and $N$ is the largest integer less than or equal to $-\left(\frac{1}{2} + \frac{2Z}{\lambda}\right)$. This means that for a bound state with an energy $E$ and angular momentum $\ell$ to occur, the coupling parameter of the inverse square component of the potential (6b) must assume one of the discrete values shown above (the PPS). For a given negative energy $E$, physical parameters $\{\ell, Z\}$ and several values of $k$, figure 1 shows a plot of the potential function (6b) and the corresponding bound state wavefunction.

It is easy to convert the PPS formula (12) into an energy spectrum formula by using Eq. (6a), which gives $\lambda$ in terms of $E$. Doing so, will give the following energy spectrum formula

$$E_k = -\frac{Z^2}{2} \bigg/ \left[k + \frac{1}{2} + \sqrt{\beta + \left(\ell + \frac{1}{2}\right)^2}\right]^2, \tag{13}$$

which is the known energy spectrum for the Kratzer potential (6b) [11]. Note that the condition $N \leq -\left(\frac{1}{2} + \frac{2Z}{\lambda}\right)$ allows the non-negative integer $k$ to assume all values from zero to infinity. Now, the wavefunctions of the bound states corresponding to the energy spectrum (13) are obtained as follows. For any choice of an energy level $k$, we use Eq. (6a) to write $\lambda_k = \sqrt{-8E_k}$. Substituting this value in the basis elements (1), we obtain

$$\phi_n^k(r) = (\lambda_k r)^{1+\nu/2} e^{-\lambda_k r/2} L_n^\nu(\lambda_k r). \tag{14}$$

The $k^{\text{th}}$ bound state wavefunction becomes $\psi_k(r) = f_0 \sum_n P_n \phi_n^k(r)$, where $\{P_n\}$ are the continuous dual Hahn polynomials $S_n(z^2; a, b, c)$ that solve the three-term recursion relation (9) with $\alpha = 2Z/\lambda_k$.

## 3. The generalized Morse potential

As a second example, we consider a one-dimensional problem on the real line and take the elements of the complete basis set as

$$\phi_n(y) = y^\mu e^{-y/2} L_n^\nu(y), \tag{15}$$

where $y(x) = e^{-\lambda x}$ is a coordinate transformation that takes the original real line, $x \in \mathbb{R}$, to $y \in [0, \infty[$. For a diagonal overlap matrix $\Omega$, the TRA solution has already been worked out where the corresponding problem turned out to be the conventional Morse potential [19]. Here, we consider the case where $\Omega$ is non-tridiagonal. In the atomic units, the action of the 1D Schrödinger wave operator on the basis element (15) reads as follows

$$\begin{aligned}-\frac{2}{\lambda^2} J |\phi_n\rangle &= \frac{1}{\lambda^2}\left[\frac{d^2}{dx^2} - 2V(x) + 2E\right] y^\mu e^{-y/2} L_n^\nu(y) = \left[y^2 \frac{d^2}{dy^2} + y\frac{d}{dy} - \frac{2V(y)}{\lambda^2} + \frac{2E}{\lambda^2}\right] y^\mu e^{-y/2} L_n^\nu(y) \\ &= y^\mu e^{-y/2}\left[y^2 \frac{d^2}{dy^2} + y(2\mu+1-y)\frac{d}{dy} + \mu^2 - \left(\mu+\frac{1}{2}\right)y + \frac{y^2}{4} - \frac{2V(y)}{\lambda^2} + \frac{2E}{\lambda^2}\right] L_n^\nu(y)\end{aligned} \tag{16}$$

Using the differential equation and differential relation of the Laguerre polynomial, this equation becomes

$$-\frac{2}{\lambda^2} J |\phi_n\rangle = y^\mu e^{-y/2} \left\{ \left[ (2\mu - \nu)n + \mu^2 - \left(n + \mu + \tfrac{1}{2}\right) y + \frac{y^2}{4} + \frac{2}{\lambda^2}(E - V) \right] L_n^\nu - (2\mu - \nu)(n + \nu) L_{n-1}^\nu \right\} \quad (17)$$

Since the integration measure is $\int_{-\infty}^{+\infty} dx = \lambda^{-1} \int_0^\infty y^{-1} dy$ then we can choose neither $2\mu = \nu + 1$ nor $2\mu = \nu + 2$ because they produce diagonal and tridiagonal basis overlap matrix $\Omega$, respectively. On the other hand, the choice $2\mu = \nu$ makes $\Omega$ a full matrix (non-zeros everywhere) and results in the following action of the wave operator one the basis

$$-\frac{2}{\lambda^2} J |\phi_n\rangle = y^{\frac{\nu}{2}+1} e^{-y/2} \left\{ \frac{y}{4} - \frac{1}{2}(2n + \nu + 1) + y^{-1} \left[ \frac{\nu^2}{4} + \frac{2}{\lambda^2}(E - V) \right] \right\} L_n^\nu. \quad (18)$$

The recursion relation and orthogonality of the Laguerre polynomials [8-10] show that the matrix representation of the wave operator $\langle \phi_m | J | \phi_n \rangle$ is tridiagonal if and only if the terms inside the curly brackets become a linear function in $y$. Therefore, we make the following choice of energy and potential function

$$(\lambda \nu)^2 = -8E, \quad (19a)$$

$$V(y) = \frac{\lambda^2}{2} y (\alpha y + \beta) = \frac{\lambda^2}{2} \left( \alpha e^{-2\lambda x} + \beta e^{-\lambda x} \right), \quad (19b)$$

where $\alpha$ and $\beta$ are arbitrary dimensionless real parameters. This is the generalized Morse potential. The conventional version of the potential corresponds to $\alpha = \tfrac{1}{4}$ [19]. Reality of $\lambda$ in (19a) dictates that the energy must be negative corresponding to bound states. Inserting (19) in Eq. (18) while setting $2\mu = \nu$ and using the three-term recursion relation of the Laguerre polynomials we obtain

$$-\frac{2}{\lambda^2} J |\phi_n\rangle = y^{\frac{\nu}{2}+1} e^{-y/2} \left\{ -\left[ (2n + \nu + 1)\left(\alpha + \tfrac{1}{4}\right) + \beta \right] L_n^\nu + \left(\alpha - \tfrac{1}{4}\right) \left[ (n + \nu) L_{n-1}^\nu + (n+1) L_{n+1}^\nu \right] \right\}. \quad (20)$$

Substituting $|\psi\rangle = \sum_n f_n |\phi_n\rangle$ in the wave equation $J|\psi\rangle = 0$ and using the above action of the wave operator on the basis, we obtain the following three-term recursion relation for the expansion coefficients

$$\left[ (2n + \nu + 1) \frac{4\alpha + 1}{4\alpha - 1} + \frac{4\beta}{4\alpha - 1} \right] P_n = (n + \nu + 1) P_{n+1} + n P_{n-1}, \quad (21)$$

where $f_n = f_0 P_n$. We compare this recursion relation to that of the Meixner polynomial $M_n(k; \gamma, c)$, which is defined in terms of the hypergeometric function as shown by Eq. (9.10.1) in Ref. [12]. If we define $\tilde{M}_n(k; \gamma, c) = c^{n/2} M_n(k; \gamma, c)$ then the recursion relation, which is given by Eq. (9.10.3) in [12], becomes

$$\left( \sqrt{c} - \frac{1}{\sqrt{c}} \right) k \tilde{M}_n = -\left[ n \left( \sqrt{c} + \frac{1}{\sqrt{c}} \right) + \gamma \sqrt{c} \right] \tilde{M}_n + (n + \gamma) \tilde{M}_{n+1} + n \tilde{M}_{n-1}, \quad (22)$$

where $\gamma > 0$, $1 > c > 0$ and $k = 0,1,2,...$ Comparing (21) with (22) gives $P_n = \tilde{M}_n(k;\gamma,c)$ and results in the following relation among the polynomial parameters and the physical parameters

$$\gamma = \nu + 1, \quad \sqrt{c} = \frac{4\alpha - 4\sqrt{\alpha} + 1}{4\alpha - 1}. \tag{23}$$

$$\beta = -\sqrt{\alpha}(2k + \nu + 1). \tag{24}$$

This dictates that $\alpha > 1/4$ and $\beta < 0$. Thus, the 1D system whose potential function is given by Eq. (19b) with the parameter $\beta$ having one of these negative discrete values (the PPS) for a given $k$ will be endowed with a bound state of energy $E$. The corresponding wavefunction is $|\psi\rangle = f_0 \sum_n P_n |\phi_n\rangle$, where $(f_0)^2$ is the positive definite weight function, which is proportional to $(\nu+1)_k c^k (1-c)^{\nu+1}/k!$ (see Eq. (9.10.2) in Ref. [12]). Using the asymptotics of $\tilde{M}_n(k;\gamma,c)$ one can show that this wavefunction vanishes for $|x| \to \infty$ as expected for a bound state. For a given negative energy $E$, physical parameters $\{\alpha, \lambda\}$ and for several values of $k$, figure 2 shows plots of the potential function (19b) and the corresponding bound states wavefunctions.

Substituting the value of $\nu$ in terms of the energy from Eq. (19a), the PPS formula (24) could be inverted to give the following energy spectrum formula

$$E_k = -\frac{\lambda^2}{8}\left[2k + 1 + (\beta/\sqrt{\alpha})\right]^2. \tag{25}$$

Imposing the condition that $\nu > -1$ and substituting $\nu = \frac{2}{\lambda}\sqrt{-2E}$ from Eq. (19a) into this energy spectrum formula dictates that $k \leq -\frac{1}{2}(1 + \beta/\sqrt{\alpha})$. Now, the wavefunctions of the bound states corresponding to the energy spectrum (25) are obtained as follows. For any choice of an energy level $k$ from this finite set, we use Eq. (19a) to write $\nu_k = \lambda^{-1}\sqrt{-8E_k}$. Substituting this value in the basis element (15), we obtain

$$\phi_n^k(y) = y^{\frac{1}{2}\nu_k} e^{-y/2} L_n^{\nu_k}(y). \tag{26}$$

The $k^{\text{th}}$ bound state wavefunction becomes $\psi_k(x) = f_0 \sum_n P_n \phi_n^k(e^{-\lambda x})$, where $\{P_n\}$ are the modified Meixner polynomials $\tilde{M}_n(k;\gamma,c)$ that solve the three-term recursion relation (22) with $\nu \mapsto \nu_k$.

## 4. The generalized Rosen-Morse potential

As a third and final example, we consider a one-dimensional problem on the real line and take the following as an element of the basis

$$\phi_n(y) = (1-y)^\alpha (1+y)^\beta P_n^{(\mu,\nu)}(y), \tag{27}$$

where $y(x) = \tanh(\lambda x)$, which transforms the original real line, $x \in \mathbb{R}$, to $y \in [-1, +1]$ and $P_n^{(\mu,\nu)}(y)$ is the Jacobi polynomial. The dimensionless real parameters $\mu$ and $\nu$ are required to be greater than $-1$. The details of the calculation for this case is relegated to Appendix B. The action of the wave operator $J$ on this basis is given by Eq. (B6). There are three scenarios where the basis overlap matrix $\Omega$ is non-tridiagonal while the matrix representation of $J$ maintains its tridiagonal structure. These scenarios are listed in equations (B10). The two cases that correspond to (B10b) and (B10c) are physically equivalent to each other. We specialize to the case (B10a) where $2\alpha = \mu$, $2\beta = \nu$ and the potential function is given by Eq. (B11a). The potential component, $V_0 \tanh(\lambda x) + \dfrac{\lambda^2 B/2}{\cosh^2(\lambda x)}$ is the 1D hyperbolic Rosen-Morse potential whose exact solution is known. On the other hand, the potential component $\dfrac{\lambda^2}{2} \dfrac{B + A \tanh(\lambda x)}{\cosh^2(\lambda x)}$ is the 1D hyperbolic pulse (single wave) potential, which is solvable only by using the TRA [20]. Considering the special case $\mu = \nu$ (i.e., $V_0 = 0$) that forces the potential to vanish at infinity, makes the potential function and basis parameter as follows

$$V(x) = \frac{\lambda^2}{2} \frac{B + A \tanh(\lambda x)}{\cosh^2(\lambda x)}, \tag{28a}$$

$$(\lambda \mu)^2 = -2E. \tag{28b}$$

Moreover, the basis (27) becomes

$$\phi_n(y) = \frac{(\mu+1)_n}{(2\mu+1)_n} (1-y^2)^{\mu/2} C_n^{(\mu+\frac{1}{2})}(y), \tag{29}$$

where $C_n^{(\mu+\frac{1}{2})}(y)$ is the ultra-spherical (Gegenbauer) polynomial and $(a)_n = a(a+1)(a+2)...(a+n-1) = \frac{\Gamma(n+a)}{\Gamma(a)}$. Assuming that $|A| \geq |B|$, the analysis in Appendix B shows that the bound state wavefunction at the negative energy $E$ is

$$\psi(x) = \frac{2f_0}{2\mu+1} [\cosh(\lambda x)]^{-\mu} \sum_{n=0}^{\infty} \frac{(1)_n (n+\mu+\frac{1}{2})}{(\mu+1)_n} H_n^{(\mu,\mu)}(z; \alpha, \theta) C_n^{(\mu+\frac{1}{2})}(\tanh(\lambda x)). \tag{30}$$

where $H_n^{(\mu,\nu)}(z; \alpha, \theta)$ is the orthogonal polynomial defined in section 3 of Ref. [21], $\cos \theta = -B/A$, $0 < \theta < \pi$, $z^{-1} = A\sqrt{1-(B/A)^2}$ and $\alpha^2 = \frac{1}{4}$. The parameter $A$ is discretized by its relation to the infinite discrete spectrum $\{z_k\}$ of $H_n^{(\mu,\nu)}(z_k; \alpha, \theta)$ via the relation $z_k^{-1} = A_k \sin \theta$ for $k = 0, 1, 2,...$ This discrete set of values $\{A_k\}$ for a given energy and $\theta$ constitutes the PPS for this problem. However, due to the absence of an analytic formula for the spectrum $\{z_k\}$, we resort to numerical means as explained at the end of Appendix B. Accordingly, we plot in figure 3 the bound states wavefunction for a given energy, parameters $\{\lambda, \theta\}$ and for several values of $k$.

For a given angle $\theta$ (i.e., parameter ratio $B/A$), we plot in figure 4 the lowest part of the PPS (for the parameter $A$) as a function of the energy. Drawing a vertical line at an energy $-E$ will intersect the curves at the PPS corresponding to that energy. On the other hand, drawing a horizontal line in the figure at any value of the potential parameter $A$ will intersect the curves at the energy spectrum corresponding to the potential parameters $A$ and $B$. The figure also shows that the problem has a finite number of bound states. Figure 5 is a plot of the energy spectrum (in atomic units) as function of the parameter ratio $(B/A)$ within the range $[-1,+1]$ and for a fixed value of the parameter $A$. Note that there is a maximum value of $(B/A)$ beyond which the potential cannot support bound states. Table 1 is a list of the finite energy spectrum (in units of $-\lambda^2$) obtained numerically for given values of the potential parameters $A$ and $B$ and for different basis sizes $N$. Stability and rapid convergence of the calculation is evident for the chosen accuracy. The bound state wavefunction corresponding to any of the energy eigenvalues in the table, $-E_k$, is obtained as the sum (30) with $\mu_k = \lambda^{-1}\sqrt{-2E_k}$ from Eq. (28b).

## 5. Conclusion and discussion

In this article, we used the Tridiagonal Representation Approach to show that if the exact solution of the wave equation is known for a single bound state at a given non-zero energy, then we could find all bound state solutions (energy spectrum and associated wavefunctions). This was accomplished by making a systematic study (with examples) of the "Potential Parameter Spectrum (PPS)", which is an infinite or finite discrete set of values of one of the potential parameters corresponding to the selected bound state energy. Inverting the relationship between the PPS and the bound state energy, we could obtain the full energy spectrum and corresponding bound states wavefunctions. This is accomplished either analytically or numerically if the spectrum formula of the associated polynomial is not known analytically.

The PPS concept could be simplified and explained in simple mathematical terms using the theory of linear ordinary differential equations as follows. A proper coordinate transformation $x \mapsto y(x)$ maps a typical wave equation in quantum mechanics,

$$\left[-\frac{1}{2}\frac{d^2}{dx^2}+V(x)-E\right]\psi(x)=0, \tag{31}$$

into the following second order linear ordinary differential equation

$$\left[a(y)\frac{d^2}{dy^2}+b(y)\frac{d}{dy}+c(y)-Eg(y)-\rho\right]\psi(y)=0. \tag{32}$$

The set of functions $\{a,b,c,g\}$ depend on $y(x)$ and $V(x)$ whereas $\rho$ is one of the potential parameters corresponding to the potential term $-\rho/g(y)$. Thus, $\rho$ becomes an eigenvalue of the second order differential operator, which is either continuous or discrete. Now, if $y(x)$ is an entire function over the whole range of $x$ and if the original equation (31) has a discrete set of solutions then $\rho$ will become a discrete set of eigenvalues of the differential operator in Eq. (32). The PPS is simply this discrete set $\{\rho_k\}$, which is associated with the given energy $E$.

## Acknowledgements

This work is supported by the Saudi Center for Theoretical Physics. Partial support by King Fahd University of Petroleum and Minerals to the Theoretical Physics Research Group under research project RG181001 is highly appreciated.

## Appendix A: The Continuous Dual Hahn Polynomial

The version of this polynomial of interest to our work could be written as follows (see Eq. 9.3.1 on page 196 of Ref. [12])

$$S_n(z^2;a,b,c) = {}_3F_2\left(\begin{matrix}-n,a+iz,a-iz\\a+b,a+c\end{matrix}\Big|1\right),\tag{A1}$$

where ${}_3F_2\left(\begin{matrix}a,b,c\\d,e\end{matrix}\Big|z\right) = \sum_{n=0}^{\infty}\frac{(a)_n(b)_n(c)_n}{(d)_n(e)_n}\frac{z^n}{n!}$ is the generalized hypergeometric function, $(a)_n = a(a+1)(a+2)...(a+n-1) = \frac{\Gamma(n+a)}{\Gamma(a)}$, $z > 0$ and $\text{Re}(a,b,c) > 0$ with non-real parameters occurring in conjugate pairs. This is a polynomial in $z^2$ which is orthogonal with respect to the measure $\rho(z;a,b,x)dz$ where the weight function reads as follows

$$\rho(z;a,b,c) = \frac{1}{2\pi}\frac{|\Gamma(a+iz)\Gamma(b+iz)\Gamma(c+iz)|^2}{|\Gamma(2iz)|^2}.\tag{A2}$$

That is, $\int_0^{\infty}S_n(z^2;a,b,c)S_m(z^2;a,b,c)\rho(z;a,b,c)dz = \xi_n\delta_{n,m}$ with $\xi_n > 0$ (see Eq. 9.3.2 on page 196 of Ref. [12]). However, if the parameters are such that $a < 0$ and $a+b$, $a+c$ are positive or a pair of complex conjugates with positive real parts, then the polynomial will have a continuum spectrum as well as a finite size discrete spectrum. In that case, the polynomial satisfies a generalized orthogonality relation (see Eq. 9.3.3 on page 197 of Ref. [12]) that contains a continuous integral and a finite sum. The polynomial satisfies the following three-term recursion relation (see Eq. 9.3.4 on page 197 of Ref. [12])

$$\begin{aligned}z^2S_n(z^2;a,b,c) &= \left[(n+a+b)(n+a+c)+n(n+b+c-1)-a^2\right]S_n(z^2;a,b,c)\\&-n(n+b+c-1)S_{n-1}(z^2;a,b,c)-(n+a+b)(n+a+c)S_{n+1}(z^2;a,b,c)\end{aligned}\tag{A3}$$

The asymptotics ($n \to \infty$) is as follows (see, for example, the Appendix in Ref. [22] where this was derived):

$$S_n(z^2;a,b,c) \approx \frac{2\Gamma(b+c)|\Gamma(2iz)|n^{-a}}{|\Gamma(b+iz)\Gamma(c+iz)\Gamma(a+iz)|}\\\left\{\cos\left(z\ln n+\arg[\Gamma(2iz)/\Gamma(a+iz)\Gamma(b+iz)\Gamma(c+iz)]\right)+O(n^{-1})\right\}\tag{A4}$$

## Appendix B: Jacobi Basis Calculation for Section 4

In this Appendix, we show details of the calculation in the Jacobi basis (27) of section 4. Using the differential chain rule $\frac{d}{dx} = \lambda(1-y^2)\frac{d}{dy}$, the action of the wave operator on the basis element (27) in the atomic units reads as follows

$$-\frac{2}{\lambda^2}J|\phi_n\rangle = \frac{1}{\lambda^2}\left[\frac{d^2}{dx^2} - 2V(x) + 2E\right](1-y)^\alpha(1+y)^\beta P_n^{(\mu,\nu)}(y) =$$

$$(1-y)^{\alpha+1}(1+y)^{\beta+1}\left[(1-y^2)\left(\frac{d}{dy} - \frac{\alpha}{1-y} + \frac{\beta}{1+y}\right)^2 - 2y\left(\frac{d}{dy} - \frac{\alpha}{1-y} + \frac{\beta}{1+y}\right) + \frac{2}{\lambda^2}\frac{E-V}{1-y^2}\right]P_n^{(\mu,\nu)}(y) \quad (B1)$$

Using the relations $\frac{1\pm y}{1\mp y} = -1 + \frac{2}{1\mp y}$ and $\frac{y}{1\pm y} = \pm 1 + \frac{\mp 1}{1\pm y}$ and collecting terms, this equation becomes

$$-\frac{2}{\lambda^2}J|\phi_n\rangle = (1-y)^{\alpha+1}(1+y)^{\beta+1}\left\{(1-y^2)\frac{d^2}{dy^2} - 2[y + \alpha(1+y) - \beta(1-y)]\frac{d}{dy}\right.$$

$$\left. -(\alpha+\beta)^2 - \alpha - \beta + \frac{2\alpha^2}{1-y} + \frac{2\beta^2}{1+y} + \frac{2}{\lambda^2}\frac{E-V}{1-y^2}\right\}P_n^{(\mu,\nu)}(y) \quad (B2)$$

Using the differential equation of the Jacobi polynomial,

$$\left\{(1-y^2)\frac{d^2}{dy^2} - [(\mu+\nu+2)y + \mu - \nu]\frac{d}{dy} + n(n+\mu+\nu+1)\right\}P_n^{(\mu,\nu)}(y) = 0, \quad (B3)$$

maps Eq. (B2) into the following

$$-\frac{2}{\lambda^2}J|\phi_n\rangle = (1-y)^{\alpha+1}(1+y)^{\beta+1}\left[\left(\frac{\mu-2\alpha}{1-y} - \frac{\nu-2\beta}{1+y}\right)(1-y^2)\frac{d}{dy}\right.$$

$$\left. + \frac{2\alpha^2}{1-y} + \frac{2\beta^2}{1+y} - (\alpha+\beta)(\alpha+\beta+1) - n(n+\mu+\nu+1) + \frac{2}{\lambda^2}\frac{E-V}{1-y^2}\right]P_n^{(\mu,\nu)}(y) \quad (B4)$$

The differential relation of the Jacobi polynomial reads as follows [8-10]

$$(1-y^2)\frac{d}{dy}P_n^{(\mu,\nu)}(y) = -n\left(y + \frac{\nu-\mu}{2n+\mu+\nu}\right)P_n^{(\mu,\nu)}(y) + 2\frac{(n+\mu)(n+\nu)}{2n+\mu+\nu}P_{n-1}^{(\mu,\nu)}(y)$$

$$= 2(n+\mu+\nu+1)\left[\frac{(\mu-\nu)n}{(2n+\mu+\nu)(2n+\mu+\nu+2)}P_n^{(\mu,\nu)}\right. \quad (B5)$$

$$\left. + \frac{(n+\mu)(n+\nu)}{(2n+\mu+\nu)(2n+\mu+\nu+1)}P_{n-1}^{(\mu,\nu)} - \frac{n(n+1)}{(2n+\mu+\nu+1)(2n+\mu+\nu+2)}P_{n+1}^{(\mu,\nu)}\right]$$

Employing this into Eq. (B4), we obtain terms on the right side that are proportional to $P_n^{(\mu,\nu)}(y)$ and $P_{n\pm 1}^{(\mu,\nu)}(y)$ as follows:

$$-\frac{2}{\lambda^2}J|\phi_n\rangle = (1-y)^{\alpha+1}(1+y)^{\beta+1}\left[2\left(\frac{\mu-2\alpha}{1-y}-\frac{\nu-2\beta}{1+y}\right)\frac{n(n+\mu+\nu+1)(\mu-\nu)}{(2n+\mu+\nu)(2n+\mu+\nu+2)}\right.$$
$$\left.+\frac{2\alpha^2}{1-y}+\frac{2\beta^2}{1+y}-(\alpha+\beta)(\alpha+\beta+1)-n(n+\mu+\nu+1)+\frac{2}{\lambda^2}\frac{E-V}{1-y^2}\right]P_n^{(\mu,\nu)}(y)$$
$$+2(1-y)^{\alpha+1}(1+y)^{\beta+1}\left(\frac{\mu-2\alpha}{1-y}-\frac{\nu-2\beta}{1+y}\right)(n+\mu+\nu+1)\times$$
$$\left[\frac{(n+\mu)(n+\nu)}{(2n+\mu+\nu)(2n+\mu+\nu+1)}P_{n-1}^{(\mu,\nu)}(y)-\frac{n(n+1)}{(2n+\mu+\nu+1)(2n+\mu+\nu+2)}P_{n+1}^{(\mu,\nu)}(y)\right]$$
(B6)

Now, the configuration space integral measure is $\int_{-\infty}^{+\infty}dx = \lambda^{-1}\int_{-1}^{+1}\frac{dy}{1-y^2}$. Thus, the orthogonality of the Jacobi polynomials,

$$\int_{-1}^{+1}(1-y)^\mu(1+y)^\nu P_n^{(\mu,\nu)}(y)P_m^{(\mu,\nu)}(y)dy = \frac{2^{\mu+\nu+1}}{2n+\mu+\nu+1}\frac{\Gamma(n+\mu+1)\Gamma(n+\nu+1)}{\Gamma(n+1)\Gamma(n+\mu+\nu+1)}\delta_{nm},$$ (B7)

and their recursion relation,

$$yP_n^{(\mu,\nu)}(y) = \frac{\nu^2-\mu^2}{(2n+\mu+\nu)(2n+\mu+\nu+2)}P_n^{(\mu,\nu)}(y)$$
$$+\frac{2(n+\mu)(n+\nu)}{(2n+\mu+\nu)(2n+\mu+\nu+1)}P_{n-1}^{(\mu,\nu)}(y)+\frac{2(n+1)(n+\mu+\nu+1)}{(2n+\mu+\nu+1)(2n+\mu+\nu+2)}P_{n+1}^{(\mu,\nu)}(y)$$
(B8)

show that a tridiagonal overlap matrix $\langle\phi_n|\phi_m\rangle$ is obtained if one of the following three corresponding relations among the basis parameter is satisfied, in the order shown below:

$$2\alpha = \begin{cases}\mu+1\\\mu+1\\\mu+2\end{cases} \qquad 2\beta = \begin{cases}\mu+1\\\mu+2\\\mu+1\end{cases}$$ (B9)

Avoiding these parameter choices, we can still obtain a tridiagonal matrix for the wave operator in (B6) if we make one of the following choices of parameters and potential

$$2\alpha = \mu, \qquad 2\beta = \nu, \qquad \frac{2}{\lambda^2}\frac{V(y)-E}{1-y^2} = \frac{2\alpha^2}{1-y}+\frac{2\beta^2}{1+y}+Ay+B.$$ (B10a)

$$2\alpha = \mu, \qquad 2\beta = \nu+1, \qquad \frac{2}{\lambda^2}\frac{V(y)-E}{1-y} = 2\alpha^2\frac{1+y}{1-y}+Ay+B.$$ (B10b)

$$2\alpha = \mu+1, \qquad 2\beta = \nu, \qquad \frac{2}{\lambda^2}\frac{V(y)-E}{1+y} = 2\beta^2\frac{1-y}{1+y}-Ay+B.$$ (B10c)

Where $A$ and $B$ are arbitrary real parameters. The last two choices are equivalent to each other as evidenced by carrying out the exchange map $\alpha \leftrightarrow \beta$, $\mu \leftrightarrow \nu$ and $y \mapsto -y$. Here, we consider the choice (B10a) which leads to the following potential function and basis parameters

$$V(x) = V_0 \tanh(\lambda x) + \frac{\lambda^2}{2} \frac{B + A\tanh(\lambda x)}{\cosh^2(\lambda x)}, \tag{B11a}$$

$$\frac{\lambda^2}{4}\left(\mu^2 - \nu^2\right) = V_0, \tag{B11b}$$

$$\frac{\lambda^2}{4}\left(\mu^2 + \nu^2\right) = -E. \tag{B11c}$$

Reality of $\mu$ and $\nu$ in (B11c) dictates that the energy must be negative corresponding to bound states. Without the $A$ term in (B11a), this is the hyperbolic Rosen-Morse potential, which has a well-known exact solution. On the other hand, with only the $A$ and $B$ terms, it becomes the hyperbolic pulse (single wave) potential, which does not have an exact solution using the conventional methods in quantum mechanics. However, using the TRA it was successfully solved by the authors in [20]. Moreover, to make the potential vanish at infinity, we must choose $V_0 = 0$ (i.e., $\mu^2 = \nu^2$) turning it into the pure hyperbolic pulse (single wave) potential [20]. Now, the choice (B10a) reduces Eq. (B6) to the following

$$-\frac{2}{\lambda^2} J|\phi_n\rangle = -(1-y)^{\alpha+1}(1+y)^{\beta+1}\left[\frac{1}{4}(\mu+\nu)(\mu+\nu+2) + n(n+\mu+\nu+1) + Ay + B\right]P_n^{(\mu,\nu)}(y)$$
$$= -(1-y)^{\alpha+1}(1+y)^{\beta+1}\left[\left(n+\frac{\mu+\nu+1}{2}\right)^2 - \frac{1}{4} + Ay + B\right]P_n^{(\mu,\nu)}(y) \tag{B12}$$

Using the recursion relation of the Jacobi polynomial (B8) for the term $Ay$ inside the square brackets will transform this equation into

$$-\frac{2}{\lambda^2} J|\phi_n\rangle = -(1-y)^{\alpha+1}(1+y)^{\beta+1}\left[\left(n+\frac{\mu+\nu+1}{2}\right)^2 - \frac{1}{4} + AC_n + B\right]P_n^{(\mu,\nu)}$$
$$-(1-y)^{\alpha+1}(1+y)^{\beta+1} A\left(D_n P_{n-1}^{(\mu,\nu)} + G_n P_{n+1}^{(\mu,\nu)}\right) \tag{B13}$$

Where the Jacobi polynomial recursion parameters $\{C_n, D_n, G_n\}$ are defined by rewriting (B8) as $y P_n^{(\mu,\nu)} = C_n P_n^{(\mu,\nu)} + D_n P_{n-1}^{(\mu,\nu)} + G_n P_{n+1}^{(\mu,\nu)}$. Using (B13) in the matrix wave equation $J|\psi\rangle = \sum_n f_n J|\phi_n\rangle = 0$ results in the following three-term recursion relation for the expansion coefficients

$$-\frac{B}{A} f_n = \left\{\left[\left(n+\frac{\mu+\nu+1}{2}\right)^2 - \frac{1}{4}\right]\frac{1}{A} + C_n\right\} f_n + G_{n-1} f_{n-1} + D_{n+1} f_{n+1} = 0. \tag{B14}$$

We define $\tilde{f}_n = R_n f_n$, where $R_{n+1} = \frac{(n+\mu+1)(n+\nu+1)(2n+\mu+\nu+1)}{(n+1)(n+\mu+\nu+1)(2n+\mu+\nu+3)} R_n$ and $R_0 = 1$. Then, we compare the resulting three-term recursion relation to that of the orthogonal polynomial $H_n^{(\mu,\nu)}(z;\alpha,\theta)$ defined in section 3 of Ref. [21]. We conclude the following:

1. If $|A| \geq |B|$ then $\tilde{f}_n = H_n^{(\mu,\nu)}(z;\alpha,\theta)$, where $\cos\theta = -B/A$, $0 < \theta < \pi$, $\alpha^2 = \frac{1}{4}$ and $z^{-1} = A\sqrt{1-(B/A)^2}$.

2. However, if $|A|<|B|$ then $\tilde{f}_n = H_n^{(\mu,\nu)}(-iz;\alpha,i\theta)$ if $A$ and $B$ have different signs. Otherwise, $\tilde{f}_n = (-1)^n H_n^{(\nu,\mu)}(-iz;\alpha,i\theta)$ where $\cosh\theta = |B/A|$, $z^{-1} = A\sqrt{(B/A)^2-1}$ and $\alpha^2 = \tfrac{1}{4}$.

Unfortunately, the spectrum formula for $H_n^{(\mu,\nu)}(z;\alpha,\theta)$ and $H_n^{(\nu,\mu)}(-iz;\alpha,i\theta)$ are not yet known. This is still an open problem in orthogonal polynomials [21,23,24]. Nonetheless, it is conjectured that $H_n^{(\mu,\nu)}(z;\alpha,\theta)$ is a pure discrete polynomial with an infinite spectrum $\{z_k\}$ [21,23]. Consequently, we resort to numerical means for calculating this spectrum, which is needed for choosing the proper set of values for the coupling parameter $A$ for a given energy and ratio $(B/A)$ (i.e., the PPS). This PPS set should be identical to that which is obtained from the spectral relation $z_k^{-1} = A\sqrt{1-(B/A)^2}$ or $z_k^{-1} = A\sqrt{(B/A)^2-1}$. For numerical stability, we choose to work with a symmetrized version of the recursion (B14) that reads

$$\frac{-1}{A}\bar{f}_n = \left[\frac{C_n + B/A}{Q_n}\right]\bar{f}_n + \sqrt{\frac{G_{n-1}D_n}{Q_{n-1}Q_n}}\bar{f}_{n-1} + \sqrt{\frac{G_n D_{n+1}}{Q_n Q_{n+1}}}\bar{f}_{n+1}, \tag{B15}$$

where $Q_n = \left(n+\tfrac{\mu+\nu+1}{2}\right)^2 - \tfrac{1}{4}$ and $\bar{f}_n = F_n f_n$ with $F_{n+1} = F_n \sqrt{Q_{n+1}D_{n+1}/Q_n G_n}$ and $F_0 = 1$.

## Figure Captions:

**Fig. 1**: The potential function with the orbital part (right column) in atomic units and the un-normalized bound state wavefunctions (left column) for the problem of Sec. 2. The horizontal dashed line is the fixed energy of all bound states at $E = -1/2$ (in atomic units). The potential parameter $\beta$ assumes one of the values given by the PPS formula (12) with $k = 0, 1, 2, 3$ (from top row to bottom row). We took $Z = -5$ and $\ell = 1$. The horizontal axis is the radial coordinate.

**Fig. 2**: The potential function (right column) in units of $\lambda^2$ and the un-normalized bound state wavefunctions (left column) for the problem of Sec. 3. The horizontal dashed line is the fixed energy of all bound states at $E = -2$ (in units of $\lambda^2$). The potential parameter $\beta$ assumes one of the values given by the PPS formula (24) with $k = 0,1,2,3$ (from top row to bottom row). The horizontal axis is the real line and we took $\alpha = 5$.

**Fig. 3**: The potential function (right column) in units of $\lambda^2$ and the un-normalized bound state wavefunctions (left column) for the problem of Sec. 4. The horizontal dashed line is the fixed energy of all bound states at $E = -1$ (in units of $\lambda^2$). The potential parameter $A$ assumes one of the values in the PPS with $k = 0,1,2,3$ (from top row to bottom row). We took the parameter ratio $B/A = 0.7$. The PPS is obtained numerically as explained in the text and shown graphically in Fig. 4 but for $B/A = -0.7$. The horizontal axis is the real line.

**Fig. 4**: The lowest part of the PPS (for parameter $A$) for the problem of Sec. 4 as a function of the energy in the range $E \in [-10, 0]$ (in units of $\lambda^2$). We took the parameter ratio $B/A = -0.7$.

**Fig. 5**: The finite energy spectrum (in units of $\lambda^2$) for the problem of Sec. 4 as a function of the parameter ratio $(B/A)$ within the range $[-1,+1]$ and for $A = 100$.

## Table Caption:

**Table 1**: The finite energy spectrum (in units of $-\lambda^2$) for the problem of Sec. 4 obtained numerically for $A = 100$ and $B = -50$ and for different basis sizes $N$. Digits that differ from those in the ($N$ = 100) column are in bold. Stability and rapid convergence of the calculation is evident.

| $N = 15$ | $N = 20$ | $N = 30$ | $N = 50$ | $N = 100$ |
|---|---|---|---|---|
| 32.76945148102**5** | 32.7694514810**22** | 32.7694514810**20** | 32.769451481023 | 32.769451481023 |
| 23.24411172615**8** | 23.24411172615**7** | 23.24411172615**8** | 23.24411172615**8** | 23.244111726155 |
| 15.1478857968**63** | 15.147885796824 | 15.147885796824 | 15.147885796825 | 15.147885796825 |
| 8.556078496**240** | 8.556078499364 | 8.55607849936**5** | 8.556078499364 | 8.556078499364 |
| 3.599423**221309** | 3.5994238**95440** | 3.599423896**582** | 3.59942389656**4** | 3.599423896564 |
| 0.569**752207819** | 0.569839**867831** | 0.569839**127953** | 0.5698390**35162** | 0.569839032667 |

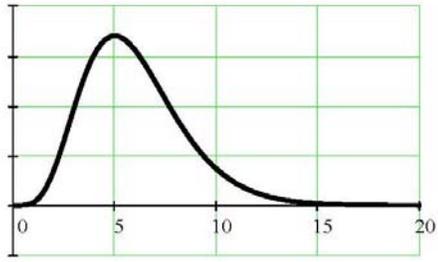 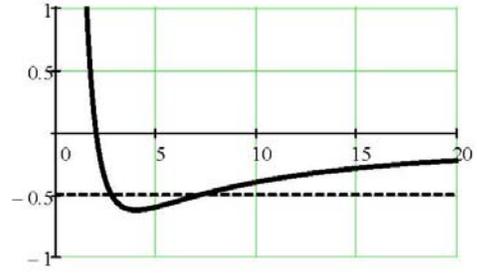
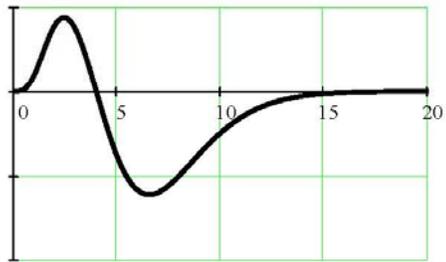 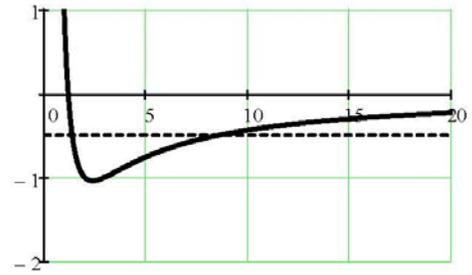
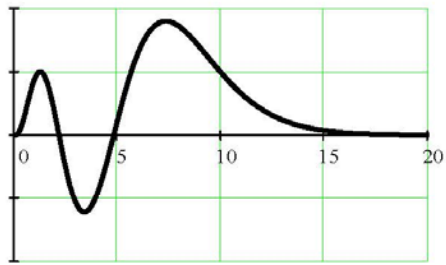 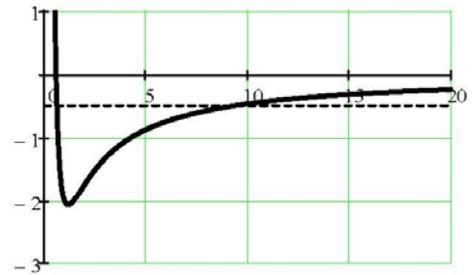
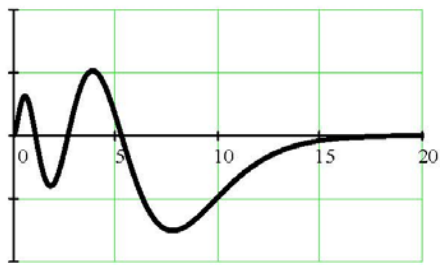 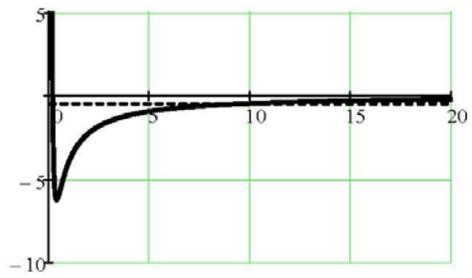

**Fig. 1**

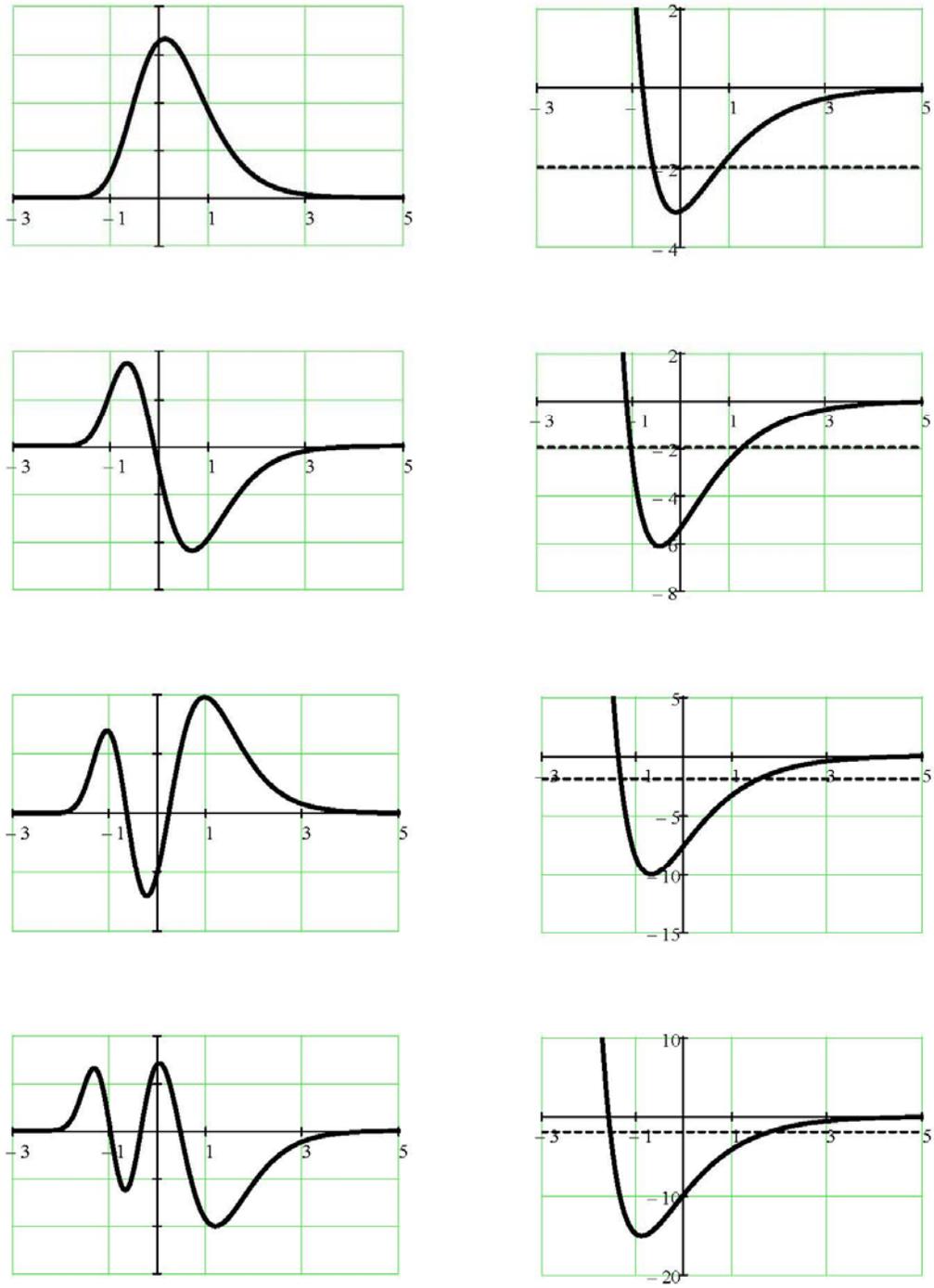

**Fig. 2**

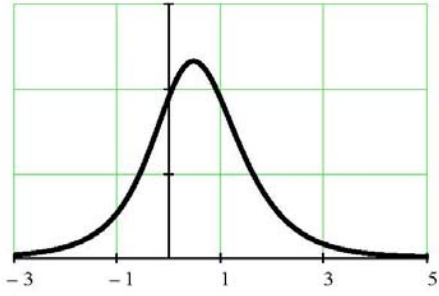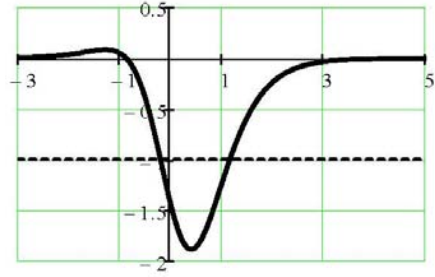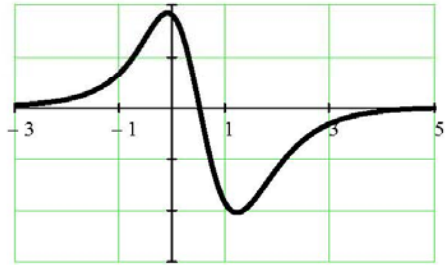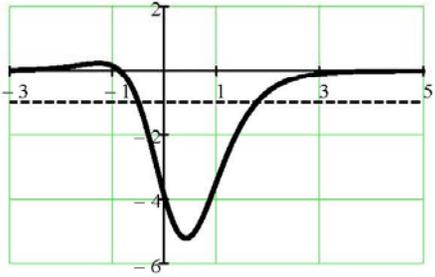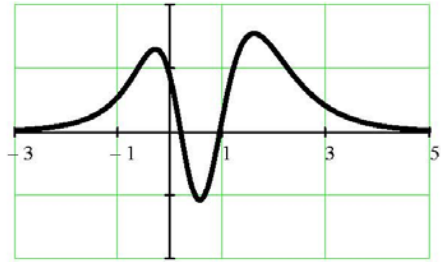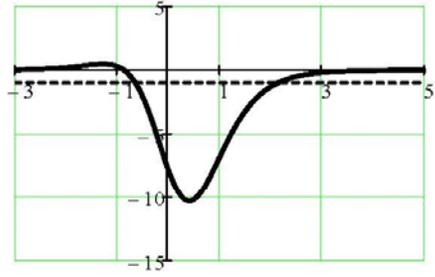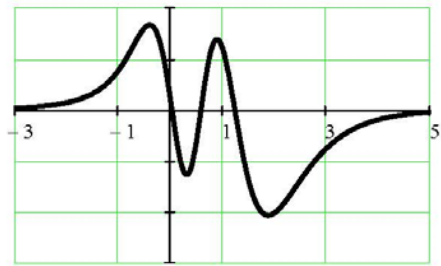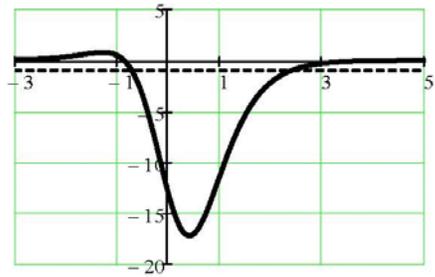

**Fig. 3**

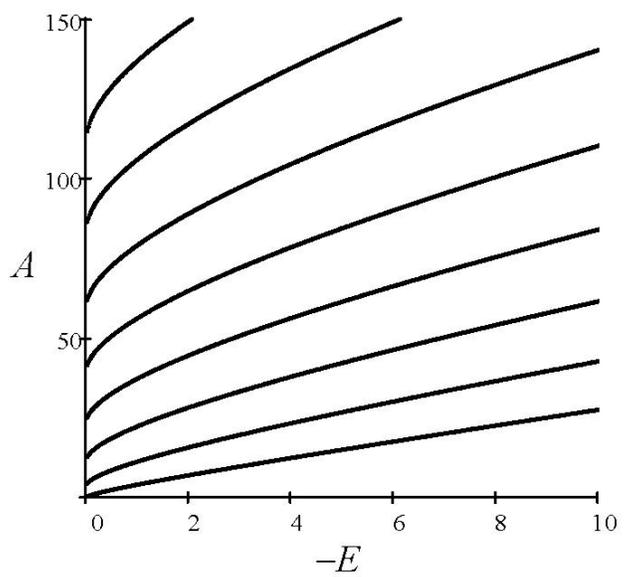

**Fig. 4**

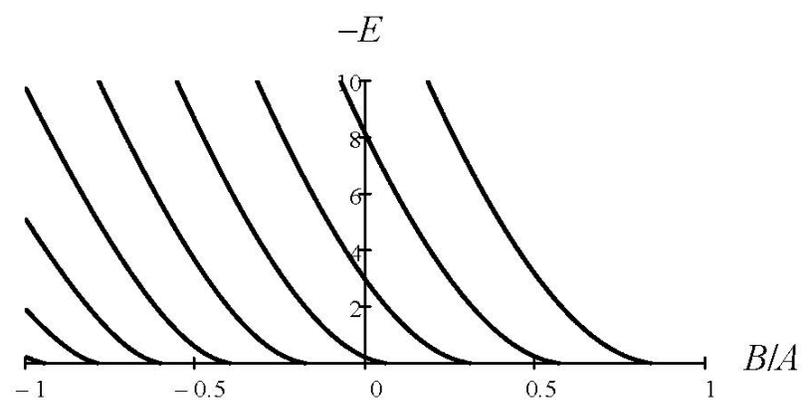

**Fig. 5**